\documentclass{aa}  

\usepackage{graphicx}
\usepackage{txfonts}
\usepackage[]{hyperref}
\usepackage{natbib}
\begin{document}

   \title{The radio continuum spectrum of Mira A and Mira B up to submillimeter wavelengths}


   \author{P. Planesas \inst{1}
          \and
          J. Alcolea \inst{1}
          \and
          R. Bachiller \inst{1}
          }

   \institute{Observatorio Astronómico Nacional (OAN-IGN),
              Alfonso XII 3, E-28014 Madrid, Spain}

   \date{Received \dots, 2015; accepted \dots, 2015}

 
  \abstract
   {}
   {We present new measurements of the flux 
densities at submillimeter wavelengths based on ALMA band 7 (338 GHz, $\lambda0.89$ mm) 
and band 9 (679 GHz, $\lambda0.44$ mm) observations to better constrain the origin 
of the continuum emission of the Mira AB 
binary system and to check its orbit.}
   {We have measured the Mira A and Mira B continuum in ALMA band 7,
with a resolution of $\sim$0\farcs31, and for the first time in ALMA band 9, 
with a resolution of $\sim0$\farcs18. 
We have resolved the binary system at both bands, 
and derived the continuum spectral index of the stars and their relative position.
We also analyzed ALMA Science Verification data obtained in bands 6 and 3. 
Measurements at centimeter wavelengths obtained by other authors have been included 
in our study of the spectral energy distribution of the Mira components.}
   {The Mira A continuum emission has a spectral index of $1.98\pm0.04$ 
extending from submillimeter down to centimeter wavelengths. 
The spectral index of the Mira B continuum emission is $1.93\pm0.06$ 
at wavelengths ranging from submillimeter to $\sim$3.1 mm, 
and a shallower spectral index of $1.22\pm0.09$ at longer wavelengths.
The high precision relative positions of the A and B components 
are shown to significantly depart from the current (preliminary) orbit 
by $\sim$14 milliarsec.}  
   {The Mira A continuum emission up to submillimeter wavelengths 
is consistent with that of a radio photosphere surrounding the evolved star 
for which models predict a spectral index close to 2. 
The Mira B continuum emission cannot be described with a single ionized 
component. An extremely compact and dense region around the star 
can produce the nearly thermal continuum measured in the range $\lambda0.4-3.1$ mm,
and an inhomogeneous, less dense, and slightly larger ionized envelope
could be responsible for the emission at longer wavelengths.
Our results illustrate the potential of ALMA for high precision astrometry of 
binary systems. We have found a significant discrepancy between the ALMA measurements
and the predicted orbit positions.}


   \maketitle
%


%

\section{Introduction}

\object{Mira A} ($o$ Ceti) is one of the best known stars in the sky. 
It is the archetype of the long-period variables (LPVs), which are cool,
pulsating, mass-losing giants on the asymptotic giant branch. 
The spectral type is M5-9IIIe+DA \citep{skiff2014}.  
Mira A is one of the nearest LVPs 
and its circumstellar envelope is one of the brightest in the 
mm-wave CO lines \citep[e.g.,][]{planesas1990a,planesas1990b}.
Its distance has been determined using different methods,
such as the Mira period-luminosity relation ($107\pm12$ pc, \citet{knapp2003}), 
the oxygen-rich Mira period-luminosity relation ($115\pm7$ pc, \citet{whitelock2008}), 
the oxygen-rich Mira period-color-luminosity relation ($105\pm7$ pc, \citet{feast1989}),
and the Hipparcos parallax ($92\pm11$ pc), which is likely the least reliable method
because the star size is three times larger than the measured parallax.
In this paper, we adopt a distance to Mira of 110 pc.

Mira A has a companion (\object{Mira B}, \object{VZ Ceti}), which is itself a variable star and 
likely to be a white dwarf \citep{sokoloski2010} 
although its nature is still controversial \citep[cf.][]{ireland2007}. This companion orbits well within 
the circumstellar envelope of Mira A and accretes gas from the giant. 
Since the visual confirmation of Mira B discovery \citep{aitken1923}, 
its position relative to the primary
has been measured to determine the orbit with little success.
The latest orbit parameters can at best be considered preliminary \citep{prieur2002}. 
The current separation on the sky is $0\farcs5$. 

In the continuum, the binary system has been resolved 
with high-resolution observations at several wavelengths: 
UV and optical with HST \citep{karovska1997}, 
X-rays with Chandra \citep{karovska2005}, 
infrared (8--18~$\mu$m) with aperture-masking \citep{ireland2007}, 
and radio, at cm wavelengths with the VLA \citep{matthews2006},
and very recently at mm wavelengths with JVLA and ALMA
\citep{matthews2015,vlemmings2015}.
In the last two papers, the radio continuum emission of Mira A 
at mm wavelengths has been resolved
and interpreted as arising in a nonuniform radio photosphere.
The Mira B continuum at mm wavelengths has been detected 
but not resolved and is more than one order of
magnitude weaker than that of Mira A.
The Mira B continuum is interpreted as arising in a circumstellar hypercompact HII region 
\citep{matthews2015} or a partially ionized wind \citep{vlemmings2015}.

In spite of the fact that previous authors have analyzed basically the same data
(ALMA Science Verification; \citeauthor{alma2015}, \citeyear{alma2015}), 
the values they obtain for the mm-wave spectral index for both stars disagree. 
\citet{vlemmings2015} 
obtain a value of $1.54\pm0.04$, and from the \citet{matthews2015}
results a value of $1.77\pm0.10$ is obtained for Mira A,
which is close to the index of 1.86 predicted by radio photosphere models \citep{reid1997}.
For the mm wavelength spectral index of Mira B, 
the results are $1.72\pm0.11$ \citep{vlemmings2015} and $1.45\pm0.10$ \citep{matthews2015},
both far from the spectral index of 2 expected for an 
optically thick ionized region.

In this paper, we present the results of radio continuum
measurements carried out with ALMA in the submillimeter bands 7 and 9, 
which are analyzed together with the high angular resolution
ALMA Science Verification data obtained in bands 3 and 6,
and complemented by cm wavelength results obtained by other authors.
Our goal is to study the cm-to-submm radio continuum 
spectrum of Mira A and B to better constrain the 
origin of the continuum emission in each star. 
The potential of ALMA for high precision astrometry 
of nearby wide binary systems is also discussed. 

\section{Observations}

We performed ALMA radio continuum observations of the Mira system 
in band 9 at 679.0 GHz on 2014 June 16
via an array configuration of 33 antennas with baselines up to 650 m.
The angular resolution was $0\farcs21\times0\farcs15$.
The continuum bandwidth was 7.95 GHz wide, 
composed of four 2 GHz wide contiguous basebands.
The system temperature was 1070 K.
Bandpass and phase calibration was performed 
pointing to the quasars J2253+1608 and J0217+0144, respectively.
Flux density calibration was performed with the quasar J2258-279, assuming a 
flux density of 0.217 Jy. 
Uranus was used as the primary flux scale calibrator
to determine the flux density of the quasar in bands 3 and 7, 
which was extrapolated to band 9 following the standard procedure
described in the ALMA Technical Handbook, where it is claimed
that an accuracy better than 15\%\ is achieved. 
As there is no guarantee that the spectral index is linear up to band 9, 
we assume a conservative accuracy of $\sim$20\%\ for the absolute flux scale.

ALMA radio continuum observations in band 7 at 338.3 GHz 
were performed on 2014 June 12, 14, and 15
using an array configuration of 34 antennas with baselines up to 650 m.
The angular resolution was $0\farcs32\times0\farcs30$.
Two 2 GHz wide basebands, centered at 332.0 and 344.6 GHz, were used 
for the continuum measurements. 
Bandpass and phase calibration was performed 
pointing to J0224+0659 (J0423-0120 on June 14) and J0217+0144, respectively.
Flux density calibration was performed with different quasars on the different dates. 
The average flux density obtained for J0217+0144, 
the phase calibrator common in all three runs, was $0.344\pm0.024$ Jy;
the uncertainty of 7\%\ reflected the different values obtained 
in the observation spectral windows and the three runs.
This proves the consistency of the ALMA flux calibration in band 7 and 
provides a measure of the uncertainty in the measured fluxes 
due to the calibration transfer to secondary calibrators, 
which is in agreement with the systematic flux uncertainty in band 7
of less than 10\%\ claimed by ALMA \citep[e.g.,][]{fomalont2014}.
On June 12 the data were taken with average atmospheric conditions 
($\sim$1.0 mm of precipitable water vapor; pwv), leading to a higher noise level than 
the data taken on June 14 and 15 at very good conditions ($\sim$0.6 mm of pwv).
We only had to perform minor manual flagging  during the calibration step.

Band 7 and band 9 observations correspond to our Cycle 1 project 2012.1.00047.S. 
The phase of Mira A was $\sim$0.1 during the observations.
Band 3 and band 6 observations correspond to the ALMA-led,
long baseline Science Verification campaign \citep{alma2015}
obtained in 2014 October and November and described elsewhere 
\citep{matthews2015,vlemmings2015}.
The observing frequency in band 3 was 94.2 GHz  
and the synthesized beam was $0\farcs070\times0\farcs060$ 
with a position angle (PA) of the major beam axis of $76\fdg6$,
counted from the north to the east.
In band 6, the observing frequency was 229.6 GHz 
and the synthesized beam was $0\farcs0344\times0\farcs0241$ with PA $= 19\fdg2$.
The Mira A phase was $\sim$0.5 during this Science Verification observing period.

\section{Data reduction}

A first look at the band 9 data revealed the strong continuum source Mira A
and an order of magnitude weaker secondary companion Mira B. 
The ALMA continuum spectrum consisted of 512 13.8 km s$^{-1}$ wide channels
where many spectral lines appear. We cleaned the maps corresponding to
the channels free of line emission 
and also excluded the end channels of each spectral window.
The data processing was performed with the CASA package 
and  we used a Briggs robust value of 0.5 for cleaning.
The rms noise in the original cleaned image was 13 mJy beam$^{-1}$, 
which is an order of magnitude higher than expected for a 219-second integration time. 
The self-calibration on the strong continuum source reduced the noise to 
1.9 mJy beam$^{-1}$ (SNR $>700$), 
after the primary beam correction was applied to produce the final image. 
The synthesized beam was $0\farcs213\times0\farcs149$ 
with a position angle of $56\fdg9$.

\begin{figure}\centering
\includegraphics[height=\vsize-60mm]{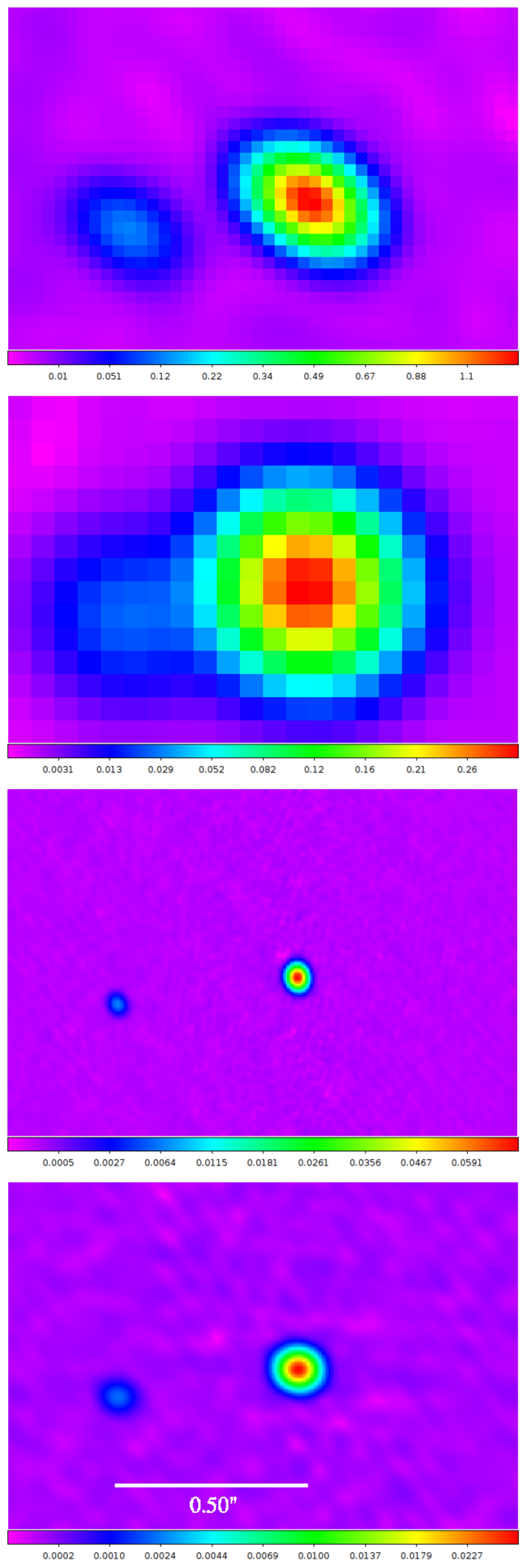} 
\caption{
  Continuum images of the Mira AB binary at the ALMA bands 9, 7, 6, and 3
  (from \emph{top} to \emph{bottom}). 
  The size of the images is $1\farcs32\times0\farcs90$.
  A square root scale has been selected for the intensity in all frames.
  The corresponding flux scale is shown at the bottom of each panel.
}\label{Fig4bands}\end{figure}

The images of Mira A and B (see Fig.~\ref{Fig4bands}, top panel)
appear clearly separated in the final band 9 image.
We performed a simultaneous two-dimensional Gaussian fit 
to determine the flux and size of each source. 
The total flux is $1460.8\pm0.5$ mJy for Mira A and $124.8\pm0.5$ mJy for Mira B.
The quoted value for the flux uncertainty is the formal error in the Gaussian fit, 
and this does not include the uncertainty in the absolute flux density calibration, 
which is much larger. 
The total flux corresponding to the emission area around each star 
is 1458 mJy for Mira A and 123 mJy for Mira B, in agreement with the previous values. 
Mira B is not resolved, within uncertainties. 
The convolved size of Mira A is slightly larger than the restored beam. 
A rough estimate of the deconvolved size of the photosphere of Mira A, 
assumed to be a uniform intensity disk, is $\sim$0$\farcs065\pm0\farcs006$. 
The higher resolution observations of the band 6 Science Verification observations 
have provided a much more precise value of 0$\farcs042\pm0\farcs002$.

Two spectral windows of 128 channels 27.2 km s$^{-1}$ wide
were used for the band 7 continuum measurements 
centered at rest frequencies of 332.0 and 344.6 GHz. 
The imaging procedure was similar to that of band 9 data: 
We only cleaned the line free channels,  
excluded the end channels of each spectral window, 
applied self-calibration, and cleaned with a Briggs robust value of 0.5.
The fitted beam used in restoration was $0\farcs316\times0\farcs304$ at PA $= 37\fdg9$.
The final image noise was 0.20 mJy beam$^{-1}$ (SNR $>1500$).

The image of Mira B in band 7 (see Fig.~\ref{Fig4bands}, second panel)
appears partially overlayed with that of Mira A, 
therefore, a simultaneous two-dimensional Gaussian fit is 
required to determine the flux of each source. Both sources are unresolved.
The total flux is $346.64\pm0.1$ mJy for Mira A and $25.5\pm0.2$ mJy for Mira B.
The same comment on the uncertainties for band 9 also applies here. 
The total fluxes, which we estimated  by defining the two emission areas with 
polygons, produced very similar results (cf. Table~\ref{table:1}).

\begin{table*}
\caption{Continuum flux density of Mira A and Mira B measured with ALMA.} 
\label{table:1}       
\centering           
\begin{tabular}{ccccccccc}       
\hline\hline         
Band & 
Frequency & 
Epoch & 
\multicolumn{3}{c}{Mira A} & &
\multicolumn{2}{c}{Mira B} \\    
\cline{4-6}\cline{8-9}
& \multicolumn{1}{c}{(GHz)}& & Phase &
\multicolumn{2}{c}{$S_\nu$ (mJy)}&& 
\multicolumn{2}{c}{$S_\nu$ (mJy)}\\    
\hline
\hline
  &        &            &      & Gaussian fit   & Polygon        && Gaussian fit   & Polygon \\
9 & 678.96 & 2014-06-16 & 0.11 & 1460.8$\pm$0.5 & 1458           && 124.8$\pm$0.5  & 123 \\
7 & 338.27 & 2014-06-12 & 0.09 &  346.4$\pm$0.1 & $\sim$347      && 25.5$\pm$0.2   & $\sim$26 \\
6 & 229.55 & 2014-10-29 & 0.51 &  152.0$\pm$0.2 & 149.7          && 11.29$\pm$0.02 & 11.0 \\
3 & $\enspace$94.19 & 2014-10-17 & 0.48 & 33.48$\pm$0.01 & 33.5  && 2.58$\pm$0.01  & 2.1 \\
\hline
\end{tabular}
\end{table*}

Our results from measurements made in mid-June 2014  
disagree with the observations of \citet{ramstedt2014} 
carried out in October 2013 with ACA and in February and May 2014 with the main array. These observations
obtained flux values roughly 30\%\ smaller than ours, 
namely 252.3 mJy for Mira A and 15.3 mJy for Mira B. 
We obtained these values  from an image made via
the emission free channels from the spectral windows
used to map the CO (3--2) emission. 
The resulting continuum image, taken with an angular resolution of $\sim$0\farcs5,
was marginally resolved, 
but the emission corresponding to the two components was not separated in the map.
The possible variability of the emission cannot explain 
the different results because the measurements with the main array were taken 
a few months apart. Moreover, \citet{ramstedt2014} 
do not mention any change in the emission in their
three epoch observations, during which the phase changed from 0.33 to 0.97.
Despite the reason for disagreement, we  later see 
that their flux values do not agree with the overall shape 
of the continuum spectrum of both stars. 

We have not calibrated the band 6 and band 3 data obtained in the Science Verification 
campaign again, as these data have been carefully calibrated
by other authors (see the comparison of different results in \citet{matthews2015}). 
Instead, we have obtained the flux density from the 
images provided by the ALMA Partnership (\citeyear{alma2015}),
both by fitting a Gaussian function to and 
by computing the flux in an area corresponding to the emission from each star. 
The results obtained with the two methods agree; cf. Table~\ref{table:1}.
The total flux values obtained for Mira A and B in band 6 
by Gaussian fitting are 152.0 mJy and 11.3 mJy, respectively,
and 33.5 mJy and 2.58 mJy in band 3.
As expected, these results agree very well with those previously obtained by 
\citet{vlemmings2015} and \citet{matthews2015}.

The results of our analysis are summarized in Table \ref{table:1}.
The values obtained with two methods (Gaussian fit, and 
total flux within a polygon surrounding the emitting area)
agree in all bands.

\section{The shape and origin of the continuum spectra}

\begin{figure}\centering
\includegraphics[width=\hsize]{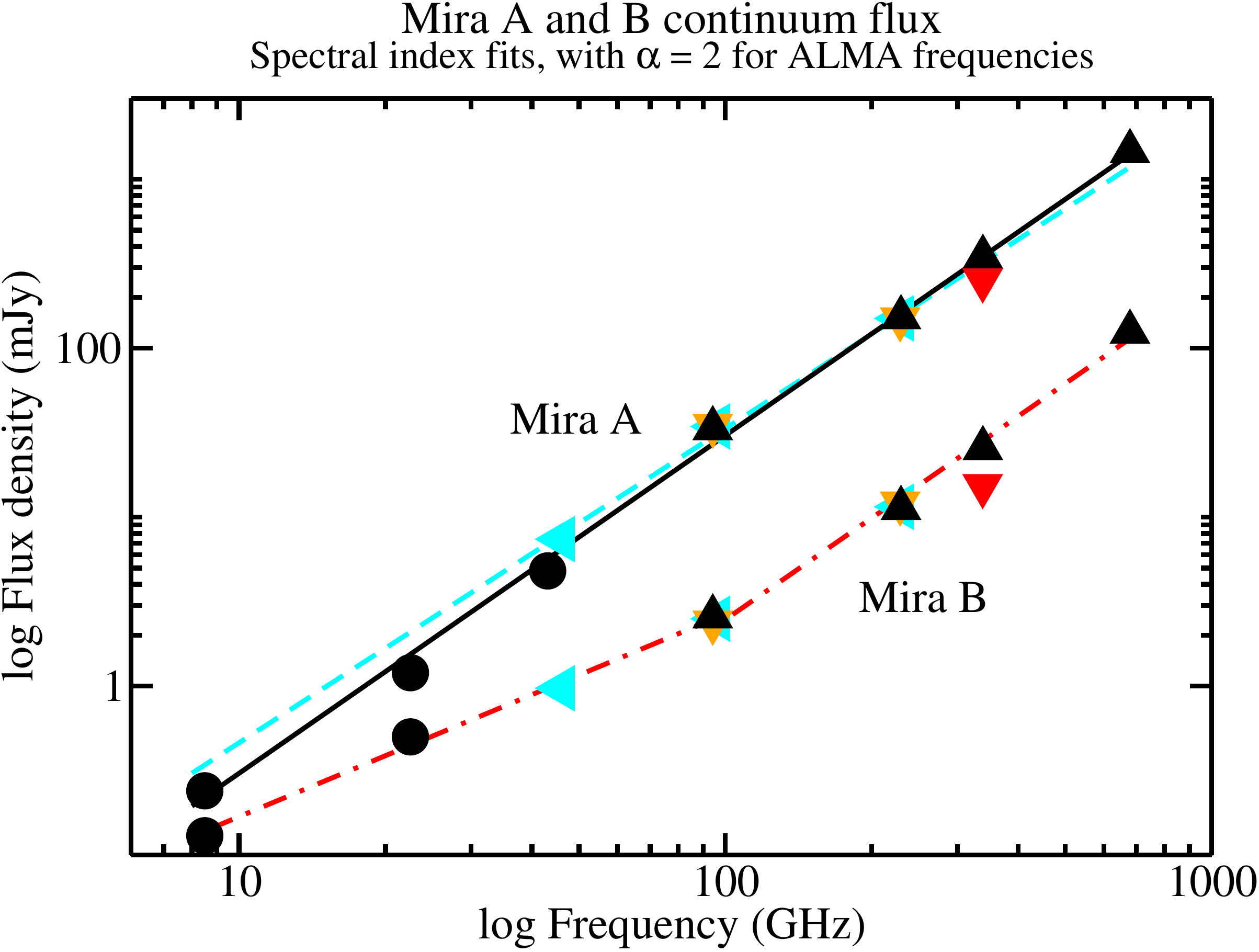}  
\caption{
  Measured values of the Mira A and Mira B continuum flux. 
  Black  upward facing triangles correspond to the ALMA measurements presented here.
  Orange downward facing triangles correspond to the results obtained from ALMA data by 
  \citet{vlemmings2015}, and red downward facing triangles to those from \citet{ramstedt2014}.
  Cyan leftward facing triangles correspond to the results obtained from ALMA and JVLA data by
  \citet{matthews2015}. 
  Black circles correspond to VLA results extracted from the papers by 
  \citet{matthews2006} and  \citet{reid2007}.
  The straight lines correspond to fits to all or to a fraction of the data 
  obtained toward each star.
}\label{FigFits}\end{figure}

The radio continuum spectra for Mira A and Mira B are shown in Fig.~\ref{FigFits},
which includes the results obtained from ALMA data presented in this paper (
black triangles), from recent papers \citep{ramstedt2014,matthews2015,vlemmings2015},
and from previous VLA data, specifically, the weighted average of measurements made 
at 8.5 and 22.5 GHz 
by \citet{matthews2006}\ and the Mira A measurements at 43.1 GHz by \citet{reid2007}.

The black line is the fit to all the Mira A measurements, which is well represented  by 
a power law with a spectral index of $\alpha=1.98\pm0.04$. 
Errors in the absolute calibration within the estimated uncertainty 
would not significantly change  this value. For example,
a reduction of 20\%\ in the band 9 flux value due to a hypothetical overestimation in the 
absolute calibration would barely reduce the spectral index to $\alpha=1.96\pm0.04$.
If a thermal spectrum is assumed, then it can be described as
$S_{\nu,A} = (31.0\pm0.6)~(\nu/100\;\mathrm{GHz})^2$ mJy.
The cyan dashed line shows the result of fitting the \citet{matthews2015} 
data with the $\nu^{1.86}$ law predicted by the radio photosphere model of
\citet{reid2007} for the centimeter wavelength range, 
which clearly deviates from the full continuum spectrum 
at the low frequency range and, possibly, at the submillimeter frequencies as well. 
On the other hand, \citet{reid2007} performed radiative transfer 
calculations for frequencies up to 10 THz ($\lambda 30\;\mu$m) 
and concluded that the spectral index remains close to 2, 
although not all the opacity contributions were considered in the model. 
In conclusion, the emission of Mira A detected with ALMA 
seems to arise mainly from the radio photosphere, 
but in the submm range additional minor contributions to the flux might exist,
such as a small contribution of warm dust in the dust shell around the star.

The continuum spectrum of Mira B cannot be described by a single power law,
as shown by the dash-dotted red lines in Fig.~\ref{FigFits}.
At frequencies below $\sim$95 GHz, the spectrum can be described as 
$S_{\nu,B} =  (2.65\pm0.06)~(\nu/100\;\mathrm{GHz})^{1.22\pm0.09}$ mJy.
At frequencies higher than $\sim$95 GHz, the spectrum is best described by a power 2 law.
In fact, the fit to the results obtained at frequencies in the range 94 to 680 GHz, 
discarding the value at 338 GHz, which  lies clearly below the straight line 
and yields a spectral index of $\alpha=1.93\pm0.06$. 
If a thermal spectrum is assumed for the high frequency range, it can be described as
$S_{\nu,B} =  (2.66\pm0.05)\;(\nu/100\;\mathrm{GHz})^2$ mJy.
\citet{matthews2015} already showed that the flux density measured in ALMA bands 6 and 7 
are well above the values that were expected from the continuum spectrum below 100 GHz,
and concluded that free-free emission from a hypercompact HII region (HCHII) 
cannot describe the full continuum spectrum. 
Our observations show that the trend extends to higher frequencies, therefore, the change in spectral index is not apparent, 
but reflects the complexity of the environment that surrounds Mira B;
this   star is embedded in the highly structured Mira A envelope \citep{ramstedt2014},
accreting matter through a circumstellar disk. 

A two-component model may be needed to explain Mira B continuum spectrum. 
The coefficient of the fit of the high frequency ($\nu>90$ GHz) part of the spectrum 
to a thermal law can be used to determine the size of the emitting region
under the assumption of optically thick emission ($\nu^2$ law) and
a typical electron temperature of $10^4$ K.
From the fit, we derive a 
half power angular size of a Gaussian distribution of $0\farcs019$, 
which corresponds to 3.1 $10^{13}$ cm at a distance of 110 pc. 
If the distribution is better represented by an uniform disk, 
then the corresponding disk size is 4.6 $10^{13}$ cm. 
These values agree with the size of $\sim4$ $10^{13}$ cm 
estimated by \citet{matthews2015} based solely in band 6 measurements,
and the size of 4.3 $10^{13}$ cm derived by \citet{vlemmings2015}.
The turnover frequency of the continuum emission is clearly larger than 700 GHz, 
therefore, the average electron density (in fact, $\sqrt{<N_e^2>}$) 
must be larger than $5\;10^8$ cm$^{-3}$. 
This value is 2-3 orders of magnitude larger than the typical density
in a HCHII region \citep{kurtz2005}. 

The continuum flux density of the lowest frequency range ($\nu<90$ GHz) 
has a  spectral index that is not as steep, although the index is not not as low as 
that of an ionized wind \citep[e.g.,][]{baez2013}. 
The continuum emission may originate in a region more extended than the 
region responsible for the thermal emission. 
A hyper-compact HII region, with a nonuniform gas
density, density gradients, or clumpy structure can produce a continuum
spectrum with a spectral index $\alpha\sim1.2$ \citep{franco2000}.
As an example, this spectral index can be produced in a nebula 
with a power-law density distribution like $N_e\propto r^{-3.2}$
\citep{olnon1975,panagia1975}. 
The turnover frequency for the ionized component responsible for the 
low frequency part of the continuum spectrum is larger than 50 GHz,
therefore, the emission measure exceeds $10^{10}$ pc cm$^{-6}$.
However, the Lyman continuum necessary to ionize a region like this 
cannot be provided solely by the white dwarf emission
owing to its low effective temperature of $<17,000$ K \citep{reimers1985}.
The accretion-power luminosity is likely to be the main source
of ionizing photons. 

We can obtain rough limit estimates of the ionized region characteristics
by assuming that the total luminosity of $10^{33}$ erg s$^{-1}$ of Mira B 
\citep{sokoloski2010} is an upper limit to the energy 
emitted in Lyman continuum photons. 
The upper limit of the ionizing photons' rate  and the lower limit of the emission measure,
taken together, imply an upper limit for the ionized region size of 1.2 $10^{14}$ cm
and a lower limit for the density of 1.6 $10^7$ cm$^{-3}$.
These parameters describe a region that is slightly larger and less dense
than the inner region, where  the thermal continuum measured in the 
3 to 0.4 mm wavelength range originates.
In fact, its size is comparable to that of the accretion disk 
as determined by \citet{ireland2007}.
In spite of all the quantitative uncertainties, we conclude that a small
ionization-limited envelope with density gradients or winds, which  explain the shallow spectral index at cm-wavelength frequencies, 
surrounds an extremely compact and dense inner region 
that emits a nearly thermal spectrum.

\begin{table*}
\caption{Relative positions of the Mira binary system.} 
\label{table:2}       
\centering           
\begin{tabular}{clllrcl}       
\hline\hline         
Epoch & \multicolumn{1}{c}{$\Delta\alpha\;\cos\delta$} & \multicolumn{1}{c}{$\Delta\delta$} & 
\multicolumn{1}{c}{$\rho$} & \multicolumn{1}{c}{$\theta$} & Telescope & Reference\\
      & \multicolumn{1}{c}{\arcsec} & \multicolumn{1}{c}{\arcsec} & 
\multicolumn{1}{c}{\arcsec} & \multicolumn{1}{c}{\degr} \\
\hline
1995.9467 & $\enspace0.549\enspace\pm0.002$ & $\enspace-0.182\enspace\pm0.002$ & 0.578 & 108.3$\enspace$ & HST & \citet{karovska1997}\\
2004.1295 & $\enspace0.525\enspace\pm0.003$ & $\enspace-0.149\enspace\pm0.003$ & 0.546 & 105.8$\enspace$ & HST & \citet{ireland2007}\\
2004.1342 & $\enspace0.5281\pm0.0008$ & $\enspace-0.1465\pm0.0008$ & 0.5480 & 105.51 & HST & This paper \\ 
2007.7283 & $\enspace0.5093\pm0.0020$ & $\enspace-0.1258\pm0.0022$ & 0.5246 & 103.87 & HST & This paper \\
2014.1478 & $\enspace0.4721\pm0.0006$ & $\enspace-0.0792\pm0.0007$ & 0.4787 &  99.53 & JVLA & Matthews, priv. comm.\\
2014.4517 & $\enspace0.4687\pm0.0006$ & $\enspace-0.0750\pm0.0005$ & 0.4747 &  99.09 & ALMA band 7 & This paper \\
2014.4572 & $\enspace0.4695\pm0.0003$ & $\enspace-0.0752\pm0.0003$ & 0.4755 &  99.10 & ALMA band 9 & This paper \\
2014.8050 & $0.46620\pm0.00011$ & $-0.07119\pm0.00010$ & 0.47161 &  98.68 & ALMA band 3 & This paper \\
2014.8309 & $0.46610\pm0.00007$ & $-0.06980\pm0.00008$ & 0.47129 &  98.52 & ALMA band 6 & This paper \\
\hline
\end{tabular}\\
\noindent $\rho$ is the star separation and $\theta$ is the position angle.
The uncertainties in the offsets are the formal errors in the Gaussian fits.\\ 
\end{table*}

\begin{figure}\centering
\includegraphics[width=\hsize]{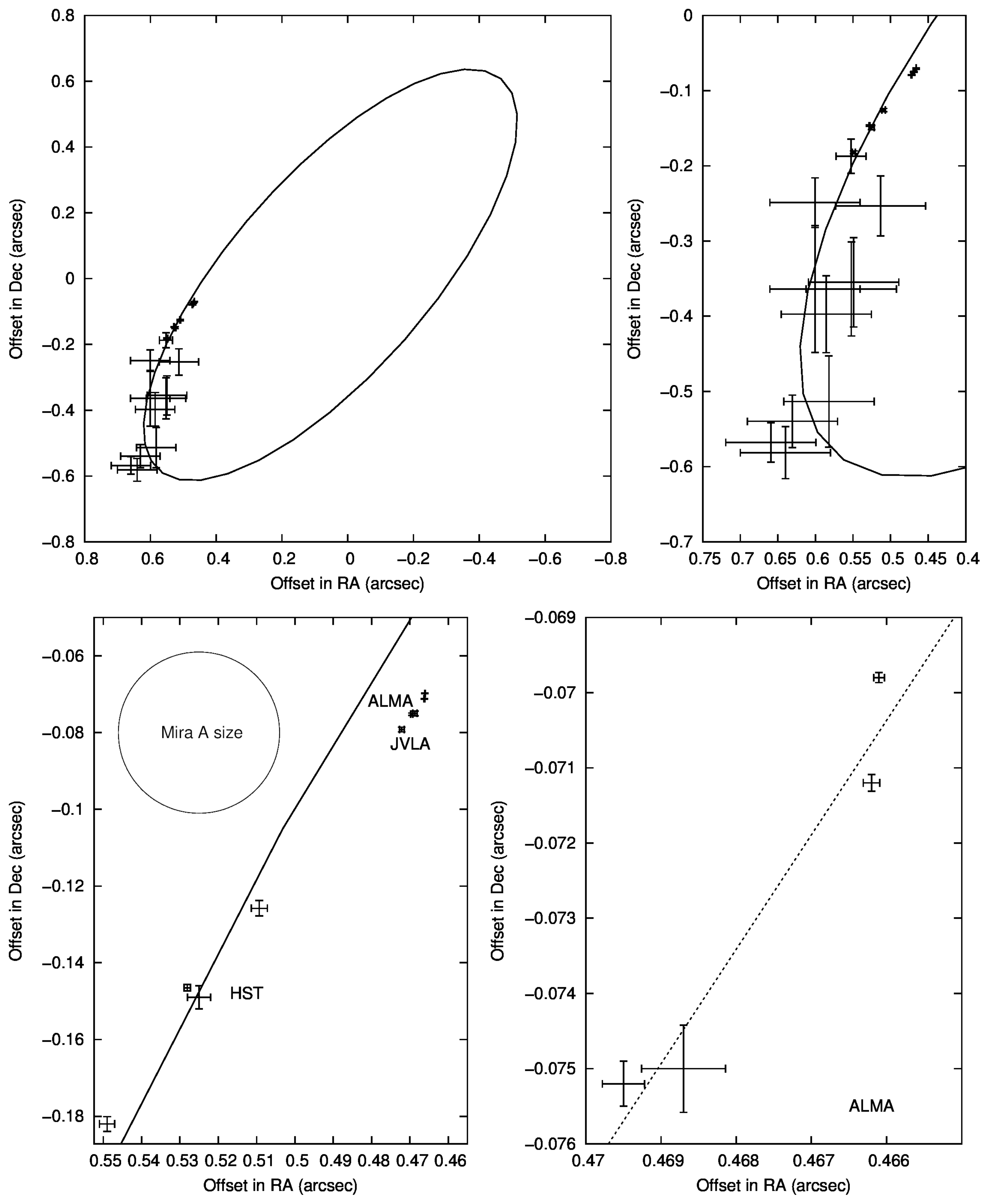}
\caption{
Relative position of the binary. In the top panels, 
the preliminary orbit determined by \citet{prieur2002} 
is overlaid on the results of optical observations done since 1923,
binned in five-year intervals. The 1-$\sigma$ dispersion 
of the values in each bin is shown with error bars. 
The HST and ALMA results of our analysis 
are included too, and they are more clearly shown in the bottom left panel
together with HST and JVLA results by other authors (cf. Table~\ref{table:2}).
The deviation from the preliminary orbit is distinctly seen. 
The circle shows the size of Mira A at 229 GHz. 
The bottom right panel shows the ALMA measurements and their formal uncertainty.
The dotted line is the result of an unweighted linear fit used to estimate the accuracy 
of the ALMA derived positions; the rms of the residuals is 0\farcs0008.
}\label{Orbit}\end{figure}

\section{Relative positions of the binary system}

The astrometric measurements carried out with ALMA can be useful to study 
the orbit of the Mira system. In fact, although the relative position of 
the two stellar components has been measured since the discovery of Mira B, 
the orbit of the Mira binary system is not yet well determined. 
The short distance between the stars
and the large brightness contrast has made the measurement 
of their relative positions rather difficult. 
Before 1980, the period of the computed orbits ranged from 14 to 841 years
\citep[cf.][]{baize1980}. In fact, an examination of the early measurements
show a large dispersion. In Fig.~\ref{Orbit} we binned the available
measurements in five-year intervals  to show the average 
position in each bin and the corresponding 1-$\sigma$ dispersion, which is
typically $0\farcs06$ in each coordinate. 
The two components were clearly separated for the first time 
in 1983 \citep{karovska1991}, using the new technique 
of speckle interferometry, which was employed thereafter. 
A further improvement in the determination of the relative position
of the binary was achieved with the Hubble Space Telescope \citep{karovska1997}.
The precision achieved in the determination of the separation of the stars 
was $\sim0\farcs003$.

The ALMA continuum measurements provide further improvement in the precision
of the determination of the relative position of the Mira components, 
down to a fraction of a milliarcsecond. 
This is the expected positional uncertainty due to the high
signal-to-noise ratio of the Mira A and B detections 
(cf. equation (1) in \citet{reid1988}) and  the rms of the residuals of a linear fit. 
This fit is a straight line, representing a very short arc of the orbit, 
to the positions determined from the ALMA measurements 
(see bottom right panel in Fig.~\ref{Orbit}).
In Table~\ref{table:2} we list the results of our analysis of the ALMA data
together with our own analysis of additional HST archival data 
(HST projects 10091 and 11224),
which are plotted in the bottom panels of Fig.~\ref{Orbit}.
The HST results lie close to the preliminary orbit, 
but this is not the case for the ALMA results, 
which clearly lie significantly apart by $\sim$0\farcs014. 
Our results for the ALMA bands 3 and 6 Science Verification data 
agree with the independent analysis made by \citet{vlemmings2015},
and with the JVLA results obtained by \citet{matthews2015}.
Therefore, we have to conclude that either 
significant structure in the radio emitting area of the binary components 
causes an apparent offset of their relative position of the centroids 
with respect to the orbit or the orbit parameters need improvement.

In fact, Mira A radio disk has been resolved using band 6 ALMA data, 
revealing the presence of a hotspot \citep{matthews2015, vlemmings2015}
slightly off-center (by $\sim$0\farcs003) of the star disk,
although these authors disagree with the value of the hotspot contribution
to the total flux. If the contribution is as large as proposed by \citet{matthews2015}, 
the location of the emission center of Mira A that would be obtained by fitting a single Gaussian,
as we have done, would be wrong by $\sim$1.4 milliarcseconds 
approximately toward the west. This would apparently shorten, by the same amount, 
the offset in right ascension (RA) between the binary components.
This offset is far too small to account for the discrepancy of $\sim$0\farcs014 
between the relative positions measured with ALMA and JVLA and the preliminary orbit. 

Therefore, we conclude that the orbit needs to be improved 
taking the new observational results into account.
However, this may not be an easy task because
the new measurements (and measurements in the decades ahead) 
correspond to a region of the orbit that is far from the ends of the apparent 
ellipse, so they are not the most favorable to compute a better orbit.

The Mira AB system is well suited to determine the mass of an AGB star,
which is usually an unknown in the studies of long period variables
and severely limits the modeling of the star evolution.
In fact, a good determination of the distance is available for Mira AB
(and a better value may be provided soon by the Gaia mission), 
therefore, the total mass of the system can be kinematically derived 
from the orbit parameters, if they are well determined. 
The mass of Mira B has been estimated to be $\sim$0.6 $M_\sun$,
both under the hypothesis that the star is a white dwarf \citep{sokoloski2010} 
or a low-mass main-sequence star \citep{ireland2007}.
Currently, with the adopted distance of 110 pc  
and the preliminary orbit computed by \citet{prieur2002}, 
we obtain $m_A+m_B \simeq 2.7$ $M_\sun$, so $m_A\sim2$ $M_\sun$,
but this value is uncertain as we have shown that the 
orbit needs to be corrected.

\section{Conclusions}

We present the results of radio continuum measurements 
carried out with ALMA in the submillimeter bands 7 and 9, 
corresponding to frequencies in the submm range 
of 338 and 679 GHz, respectively. 
We reduced the data  and  produced maps, 
after removing the numerous spectral lines that appear
in the observed spectral windows. 
High angular resolution ALMA Science Verification data obtained 
in bands 3 and 6 in the mm-wavelength range have also been reduced.
The flux determined with ALMA for both Mira components
follow a power law of spectral index $\sim$2.0. 

The Mira A continuum emission is well described by an optically thick 
thermal spectrum of spectral index $\alpha = 1.98\pm0.04$, 
which extends from cm-to-submm wavelengths (frequency from 8 to 700 GHz). 
This is in agreement with radiative transfer models, which 
conclude that the spectral index of the radio photosphere that surrounds 
the evolved star remains close to 2 for frequencies up to the THz range
\citep{reid2007}.

The Mira B continuum spectrum is described with spectral index 
$\alpha = 1.22\pm0.09$ at frequencies below $\sim$95 GHz and
$\alpha = 1.93\pm0.06$ at higher frequencies up to the submm range.
Therefore, it cannot be described with a single ionized component.
We propose a two-component model, 
consisting of an extremely compact and dense region 
around the star than can produce the nearly thermal continuum
measured at the high frequency range 
and, surrounding this region, a slightly larger and less dense, 
nonuniform ionized envelope with density gradients or winds that are
responsible for the emission at longer wavelengths.
The ionizing source is likely associated with the accretion process. 

The potential of ALMA for high precision astrometry 
of nearby binary systems has been discussed. 
In a few minutes of observing time, the positions can be determined
with great accuracy for distances up to 100 pc, even in ALMA band 9. 
The ALMA continuum measurements provide an order of magnitude 
improvement with respect to HST measurements 
in the precision of the determination of the relative 
position of the Mira components down to a tenth of a milliarcsecond.
However, because of the extremely high angular resolution of ALMA, 
which is able to resolve the structure in the stellar disks, 
some modeling may be required  to determine the 
geometrical center of the disks. 
Nevertheless, we have found a significant discrepancy, 
of the order of 0\farcs014, between the ALMA measurements 
and the predicted orbit. This discrepancy is evidence that the orbit parameters 
need to be improved taking  the high precision results
of the current observations into account. 
Better orbit parameters would help to constrain the Mira A mass
and allow us to improve the evolutionary modeling of this AGB star.

\begin{acknowledgements}
This paper makes use of the following ALMA data:
ADS/ JAO.ALMA\#2013.1.00047.S. ALMA is a partnership of ESO (representing
its member states), NSF (USA), and NINS (Japan) together with NRC
(Canada),  NSC, and ASIAA (Taiwan) in cooperation with the Republic of
Chile. The Joint ALMA Observatory is operated by ESO, AUI/NRAO, and NAOJ.
Also, some data is based on observations made with the NASA/ESA Hubble Space Telescope, 
obtained from the data archive at the Space Telescope Science Institute. 
STScI is operated by the Association of Universities for Research in Astronomy, 
Inc. under NASA contract NAS 5-26555.
This research has been partially funded by the Spanish DGCyT grants FIS2012-32096
and FIS2012-32032, 
and has received travel funding from 
the European Commission Seventh Framework Programme (FP/2007-2013) 
under grant agreement No 283393 (RadioNet3) through the RadioNet
Networking Activity MARCUs (Mobility for ALMA Regional Centre Users). 
The authors thank Edwige Chapillon
and the IRAM ARC node for their assistance in the data calibration, 
and Lynn Matthews for providing additional information on Fig. 1 
in her 2015 paper.  
\end{acknowledgements}

\end{document}